# Optimization of temperature sensitivity using the optically detected magnetic resonance spectrum of a nitrogen-vacancy center ensemble

Kan Hayashi[1,2,3], Yuichiro Matsuzaki[3], Takashi Taniguchi[4], Takaaki Shimo-Oka[1], Ippei Nakamura[2][†], Shinobu Onoda[5], Takeshi Ohshima[5], Hiroki Morishita[1], Masanori Fujiwara[1], Shiro Saito[3], Norikazu Mizuochi[1]

[1]*Instutute for Chemical Research, Kyoto University, Uji, Kyoto 611-0011, Japan*

[2]*Graduate School of Engineering Science, Osaka University, Toyonaka, Osaka 560-8531, Japan*

[3]*NTT Basic Research Laboratories, NTT Corporation, Atsugi, Kanagawa 243-0198, Japan*

[4]*National Institute for Materials Science (NIMS), Tsukuba, Ibaraki 305-0044, Japan*

[5]*National Institutes for Quantum and Radiological Science and Technology (QST), Takasaki, Gunma 370-1292, Japan*

[†]*Present address: RIKEN Center for Emergent Matter Science (CEMS), Wako, Saitama, 351-0198, Japan*

Abstract

Temperature sensing with nitrogen vacancy (NV) centers using quantum techniques is very promising and further development is expected. Recently, the optically detected magnetic resonance (ODMR) spectrum of a high-density ensemble of the NV centers was reproduced with noise parameters [inhomogeneous magnetic field, inhomogeneous strain (electric field) distribution, and homogeneous broadening] of the NV center ensemble. In this study, we use ODMR to estimate the noise parameters of the NV centers in several diamonds. These parameters strongly depend on the spin concentration. This knowledge is then applied to theoretically predict the temperature sensitivity. Using the diffraction-limited volume of 0.1 μm$^3$, which is the typical limit in confocal microscopy, the optimal sensitivity is estimated to be around 0.76 mK/$\sqrt{\text{Hz}}$ with an NV center concentration of $5.0 \times 10^{17}$/cm$^3$. This sensitivity is much higher than previously reported sensitivities, demonstrating the excellent potential of temperature sensing with NV centers.

## I. INTRODUCTION

A negatively charged nitrogen-vacancy (NV) center in diamond [1,2] is a promising system to realize practical quantum devices [3-23]. The spin state of an NV center can be initialized, manipulated, and read out at room temperature [3, 4]. Due to the excellent controllability and long coherence time [19-21], entanglement generation between NV centers has been already demonstrated [3-11]. These techniques have been applied to NV centers for quantum information proceedings [3-6], quantum hybrid device [22-24], and quantum sensing [7-18]. In particular, single NV centers have been used to detect magnetic fields [7-11], temperature [11-14], pressure [15], and electric fields [16-18] on the nanometer scale. Not only a single NV center but also a high-density ensemble of NV centers has been studied for quantum sensing applications. Because the sensitivity increases by $\sqrt{N}$ as the number of the NV centers contributed to measurements ($N$) increases, an ensemble of the NV centers is expected to be a suitable high-sensitive sensor with a spatial resolution on the submicron to millimeter scale. Due to various potential applications, the basic properties of the NV center ensemble have been intensively studied [7, 24-28].

The NV center constitutes a spin-1 system in the ground-state manifold. Figure 1 shows its energy diagram. The $|m_s = 0\rangle$ and the $|m_s = \pm 1\rangle$ levels are split by the axial zero-field splitting (ZFS) $D_g \approx 2.87$ GHz. Without an applied magnetic field, the energy eigenstates are described as $|B\rangle = \frac{1}{\sqrt{2}}(|1\rangle + |-1\rangle)$ and $|D\rangle = \frac{1}{\sqrt{2}}(|1\rangle - |-1\rangle)$, where these two states are energetically split by the strain (electric) fields. On the other hand, applying an external magnetic field $B$ larger than the strain (electric) field, the energy eigenstates are $|m_s = \pm 1\rangle$, and these states are separated by the Zeeman energy of $2g_e\mu_B B$, where $g_e = 2.003$ is the NV center $g$ factor and $\mu_B$ is the Bohr magneton. The NV center spin state can be optically initialized into $|m_s = 0\rangle$, and measuring the spin-dependent fluorescence can read the spin states. Hence, the magnetic resonance between the energy eigenstates in the ground-state manifold can be optically detected. From the resonant frequencies of the optically detected magnetic resonance (ODMR), the temperature, magnetic fields, and electric fields can be estimated.

The NV center Hamiltonian, which contains the effects of the temperature, magnetic fields, and electric field, is expressed as

$$H = D_g(T, \varepsilon_z)S_z^2 + g_e\mu_B B_z S_z + g_e\mu_B(B_x S_x + B_y S_y)$$
$$+ c_{\varepsilon_\perp}\epsilon_x(S_x^2 - S_y^2) + c_{\varepsilon_\perp}\varepsilon_y(S_x S_y + S_y S_x)$$
$$+ \lambda \cos(\varpi t) S_x + A_\parallel S_z I_z + \frac{A_\perp}{2}(S_+ I_- + S_- I_+) + P(I_z^2 - \frac{1}{3}I^2) - g_n\mu_N B_z I_z,$$

where $S$ ($I$) is a spin-1 operator of the electron (nuclear) spin. $\epsilon_{x\,(y)}$ is a term to represent strain (electric) fields along the $x$- or $y$-axis. $g_e\mu_B B_z S_z$ ($g_n\mu_N B_z I_z$) is the Zeeman term of the electron (nuclear) spin. $\lambda$ is the microwave amplitude, and $\omega$ is the microwave driving frequency. $P$ is the quadrupole splitting. $A_\parallel$ ($A_\perp$) is the parallel (perpendicular) hyperfine coupling. $c_{\varepsilon_\perp} = 2\pi \times$

170 kHz /(V) μm. $D(T, \varepsilon_z)$ is expressed as

$$D(T, \varepsilon_z) = 2\pi \times 2870.685 \text{ MHz} + C_T \Delta T + c_{\varepsilon_\parallel} \varepsilon_z,$$

where $C_T = 2\pi \times (-78.6 \pm 0.5 \text{ kHz/K})$ and $c_{\varepsilon_\parallel} = 2\pi \times 3.5 \text{ kHz /(V) μm}$. Since the equation contains terms of $g_e \mu_B (= 2\pi \times 28.7 \text{ GHz/T})$, $c_{\varepsilon_\perp}$, and $C_T$, the resonant frequencies in the ODMR depend on the magnetic field, electric field, and temperature. Here, the *x*-axis is defined as parallel to the direction of the applied microwave, while the *z*-axis is parallel to the NV center axis. It should be noted that although $c_{\varepsilon_\parallel}$ and $g_n \mu_N (= 2\pi \times 3.08 \text{ MHz/T})$ have finite contributions to the ODMR signal, their effects are much smaller. Consequently, these are ignored in this paper.

The uncertainty of an estimated parameter with a diamond-based sensor is given by

$$\partial S_{min} \sim \frac{1}{\left(\left|\frac{\partial P_{signal}}{\partial S_m}\right|_{max} \sqrt{N}\right)},$$

where $S_m$ is the amplitude of the target parameter (temperature, electric field, or magnetic field). $P_{signal}$ is the optical signal intensity from the NV center. This formula shows that the sensitivity can be improved by increasing either the gradient of the ODMR signal or the number of the NV centers. However, as the concentration of the NV centers increases, noise effects such as inhomogeneous magnetic fields, strain (electric field) variations, and homogeneous broadening tend to be larger, increasing the linewidth in the ODMR signal. This means that there is a trade-off relationship between $|\partial P_{signal}/\partial S_m|_{max}$ and $N$. Thus, it is not trivial to find the optimized concentration of the NV centers for high-performance sensing devices. This motivated us to understand the influence of the NV centers on the noise parameters.

It was previously revealed that the ODMR spectrum for a high-density ensemble of NV centers shows a sharp dip structure around 2870 MHz without an applied magnetic field. This cannot be fitted by the sum of the Lorentzian functions [22]. Zhu et al. [22, 28] successfully reproduced this sharp dip structure using a theoretical model that contained the effects of inhomogeneous magnetic fields, inhomogeneous strain (electric field) distributions, and the homogeneous width of the NV centers. Matsuzaki et al. [28] showed that the sharp dip structure is robust against inhomogeneous magnetic fields and strain (electric field) variations, but it is sensitive against the homogeneous broadening. They also introduced a schematic way to estimate the noise parameters of the NV centers from the fitting of the ODMR spectra.

In this paper, we use Matsuzaki's scheme to estimate the noise parameters of the NV centers in several diamond samples and demonstrate the dependence of the noise parameters on the spin concentration. Although there are many qualitative discussions about the dependence of the noise parameters on the spin concentration [28-31], the aim of this paper is to quantitatively analyze the dependence by comparing theory and experiment. In particular, we reveal the relationship between the NV center concentration and the strain (electric field) distribution. Such a quantitative analysis has yet to be reported.

As an application of our analysis, we discuss the sensitivity at high concentrations of the NV center ensemble as a temperature sensor. Our analysis provides a way to estimate the change in the noise parameters as a function of the NV center concentration. Then the dependence is simulated for each NV center ensemble concentration. By calculating the sensitivity at each concentration, the optimized density of the NV centers for the minimum sensitivity is estimated. We propose using the dip structure in the ODMR spectrum where the gradient around the dip structure becomes three times larger than that used for a temperature sensor in previous research. We compare the sensitivity of our scheme using the dip structure with that of the conventional scheme. Finally, we demonstrate that our scheme using an NV ensemble has the potential to provide a better temperature sensor.

This article is structured as follows. Section II describes the sample fabrication and the measurement setup. Section III details the theoretical model. Section V presents the ODMR results for several diamond samples, the noise parameter dependences, and the theoretical estimation of the temperature sensitivity of the NV centers. Section V summarizes the results.

## II. EXPERIMENTAL METHOD

Diamonds, including substitutional nitrogen (P1) centers, were synthesized by high pressure high temperature HPHT processes. The NV centers were prepared with electron irradiation and subsequent annealing at 1000 °C for 1 hour in Ar gas. Afterwards, the diamonds were cleaned in a mixture of boiling acids (1:1 sulfuric, nitric) to remove graphitic carbon and oxygen-termination at the surface. The P1 center concentrations were investigated by electron paramagnetic resonance (EPR) spectroscopy at room temperature, while the NV center concentrations were investigated by the fluorescence intensity when a 532-nm laser was applied. The diamonds were measured by ODMR spectroscopy using a lab-built confocal microscope at room temperature. It is possible to initialize and read out the NV centers by illuminating with a green laser (532 nm) pulse (100 μsec) at 50 μW.

Microwave pulses were generated via a copper wire, which was placed close to the diamond surface. This setup allowed the NV centers to be controlled. The NV centers had a Rabi frequency $\Omega = 2\pi \times 125$ kHz for a pulse duration around 4 μsec. To apply an external magnetic field, we used a permanent magnet. Suitable magnetic fields were applied on the diamonds, inducing a separation of the ODMR signals into four groups with different crystallographic axes in the diamond lattice.

## III. MODEL

We introduce a model to simulate the ODMR spectra [28]. The Hamiltonian of the NV center is described as

$$H = DS_z^2 + g_e\mu_B B_z S_z + E_1(S_x^2 - S_y^2) + E_2(S_x S_y + S_y S_x) + \lambda \cos(\varpi t) S_x,$$

where $E_1 = c_{\varepsilon_\perp}\epsilon_x$, $E_2 = c_{\varepsilon_\perp}\varepsilon_y$. The $x$ and $y$ components of the magnetic fields are negligible because the Zeeman energy is assumed to be much smaller than the zero-field splitting ($D$). Since we consider the NV center ensemble, the total Hamiltonian in a rotating frame, which is defined by $U = e^{-i\omega S_z^2 t}$, is described as

$$H \sim \hbar \sum_{k=1}^{N} \left\{ (D_k - \omega)S_{z,k}^2 + E_1^{(k)}(S_{x,k}^2 - S_{y,k}^2) + E_2^{(k)}(S_{x,k}S_{y,k} + S_{y,k}S_{x,k}) + g_e\mu_B B_z^{(k)} S_z + \frac{\lambda}{2} S_x^{(k)} \right\}$$

$$= \hbar \sum_{k=1}^{N} \left\{ (D_k - \omega)(|B\rangle_k\langle B| + |D\rangle_k\langle D|) + E_1^{(k)}(|B\rangle_k\langle B| - |D\rangle_k\langle D|) + iE_2^{(k)}(|B\rangle_k\langle D| - |D\rangle_k\langle B|) + g_e\mu_B B_z^{(k)} E_2^{(k)}(|B\rangle_k\langle D| + |D\rangle_k\langle B|) + \frac{\lambda}{2}(|0\rangle_k\langle B| + |B\rangle_k\langle 0|) \right\},$$

where $|B\rangle_k = \frac{1}{\sqrt{2}}(|1\rangle_k + |-1\rangle_k)$, $|D\rangle_k = \frac{1}{\sqrt{2}}(|1\rangle_k - |-1\rangle_k)$. By assuming that $D_k$, $B_z^{(k)}$, $E_1^{(k)}$, and $E_2^{(k)}$ are randomly distributed, the effect of the inhomogeneous broadening can be included. More specifically, the random distributions about $D_k$, $E_1^{(k)}$, and $E_2^{(k)}$ are described by a Lorentzian function. The inhomogeneous strain (electric field) distribution along the $z$-axis included in $D_k$ is set to 1/50 of the magnitude of the inhomogeneous strain (electric field) distribution along $x$- and $y$-axes because $c_{\varepsilon_\parallel}$ is 50 times smaller than $c_{\varepsilon_\perp}$ [16]. The effect of other nuclear spins ($^{15}$N, $^{13}$C) is considered as randomized magnetic fields on the electron spin. P1 centers are also the sources of the randomized magnetic fields. In this model, we include these effects by a Lorentzian probability distribution of the magnetic fields applied on the NV centers. It should be noted that the hyperfine interaction from the nitrogen nuclear spins induces an effective magnetic field on the NV centers. This leads to a random distribution of the magnetic fields ($B_z^{(k)}$). In our experimental conditions, the nitrogen nuclear spin in the NV becomes completely mixed state, which means that the state of the nitrogen nuclear spin is either |-1>, |0>, or |1> with an equally probability distribution. Depending on the state of the nitrogen nuclear spin, the effective magnetic fields on the NV centers is different. This means that we can consider the distribution of the as a Lorentzian functions, which are separated by $2\pi \times 2.3$ MHz, due to the fact that the state of the nuclear spin become almost completely mixed at a room temperature. Such a theoretical model has been used to reproduce experimental results about the ensemble of the NV centers by many researchers [22-24, 28].

To include the effect of the homogeneous broadening [28], we adopt a Lindblad type master equation, which is expressed as

$$\frac{d\rho(t)}{dt} = -\frac{i}{\hbar}[H,\rho(t)] + \sum_{k=1}^{N}\sum_{j=1}^{4} \gamma_{j,k}(2L_{j,k}\rho(t)L_{j,k}^\dagger - L_{j,k}^\dagger L_{j,k}\rho(t) - \rho(t)L_{j,k}^\dagger L_{j,k}),$$

where $\gamma_{j,k}$ ($j = 1,2,3,4$) is the decay rate and $L_{j,k}$ ($j = 1,2,3,4,$) is the Lindblad operator. In the case of the NV centers, $L_{1,k} = |B\rangle_k\langle D|$, $L_{2,k} = |D\rangle_k\langle B|$, $L_{3,k} = |0\rangle_k\langle B|$, and $L_{4,k} =$

$|0\rangle_k \langle D|$, where $\gamma_{1,k} = \gamma_{2,k} = \Gamma$ is the dephasing rate of the high-frequency magnetic field noise and $\gamma_{3,k} = \gamma_{4,k} = \Gamma'$ is the energy relaxation rate. Since the NV center has a much longer energy relaxation time than the dephasing time, as described below, the limit of the small energy relaxation rate is used. Additionally, we assume a steady state. Hence, the time derivative of $\rho(t)$ can be set as zero. Moreover, to avoid power broadening, the microwave strength $\lambda$ is considered to be much smaller than the other parameters. From these assumptions, an analytical solution of the master equations can be obtained [28].

$$P_0 = 1 - \frac{1}{N} \sum_{k=1}^{N} \text{Tr}[\rho(\infty)|0\rangle\langle 0|],$$

$$= 1 - \frac{1}{N} \left( \sum_{k=1}^{N} \left| \frac{\lambda' \left( \varpi - \varpi_d^{(k)} + i\Gamma'' \right)}{\left( \varpi - \varpi_b^{(k)} + i\Gamma'' \right)\left( \varpi - \varpi_d^{(k)} + i\Gamma'' \right) - \left( |J_k|^2 + |J_k'|^2 \right)} \right|^2 \right.$$

$$\left. + \left| \frac{\lambda'(J_k - iJ_k')}{\left( \varpi - \varpi_b^{(k)} + i\Gamma'' \right)\left( \varpi - \varpi_d^{(k)} + i\Gamma'' \right) - \left( |J_k|^2 + |J_k'|^2 \right)} \right|^2 \right),$$

where $\varpi_b^{(k)} = D_k - E_1^{(k)}$, $\varpi_d^{(k)} = D_k + E_1^{(k)}$, $J_k = g\mu_B B_z^{(k)}$, $J_k' = E_2^{(k)}$, $\Gamma'' = \Gamma + \Gamma'$, $\lambda' = \sqrt{\frac{\Gamma''}{\Gamma'}} \lambda$. We take the limit of a small energy relaxation rate $\Gamma'$, while keeping $\lambda'$ constant. It should be noted that the higher order terms of $\lambda'$ are dropped because the microwave driving is assumed to be sufficiently weak. Thus,

$$P_0 = 1 - \frac{1}{N} \left( \sum_{k=1}^{N} \left| \frac{\lambda' \left( \varpi - \varpi_d^{(k)} + i\Gamma \right)}{\left( \varpi - \varpi_b^{(k)} + i\Gamma \right)\left( \varpi - \varpi_d^{(k)} + i\Gamma \right) - \left( |J_k|^2 + |J_k'|^2 \right)} \right|^2 \right.$$

$$\left. + \left| \frac{\lambda'(J_k - iJ_k')}{\left( \varpi - \varpi_b^{(k)} + i\Gamma \right)\left( \varpi - \varpi_d^{(k)} + i\Gamma \right) - \left( |J_k|^2 + |J_k'|^2 \right)} \right|^2 \right).$$

This approach can reproduce the ODMR spectra with high NV center concentrations.

**IV. RESULTS AND DISCUSSION**

Table 1 shows P1, NV, and the spin concentrations (which are the sum of the P1 and the NV concentrations) for the four diamond samples estimated from the EPR and the fluorescence measurements. Fitting our numerical simulation to the ODMR spectra of the samples determined the noise parameters of the samples. Figures 2(a), (c), (e), (g) and 2(b), (d), (f), (h) show the ODMR simulation results without and with an external magnetic field, respectively. In the simulation, we changed the parameter $\Gamma/2\pi$ (homogeneous broadening), $dg\mu B_z/2\pi$ (inhomogeneous magnetic fields), and $dE_{1,2}/2\pi$ (inhomogeneous strain (electric field) distributions). All other parameters were fixed. The sharp dip in the ODMR spectrum strongly depends on $\Gamma/2\pi$, but the dip structure is insensitive to the change in $dg\mu B_z/2\pi$ and $dE_{1,2}/2\pi$. Consequently, $\Gamma/2\pi$ can be estimated from the shape of the sharp dip. Strictly speaking, there must be an effect of the earth magnetic fields (of around 0.045 mT) on the NV centers, even when experimentalists did not apply any external magnetic fields. However, by numerical simulations, we have checked that the effect of the earth magnetic fields is negligibly small in the ODMR, and the existence of the earth magnetic fields does not qualitatively change our results (Fig. 3).

The ODMR signal in an external magnetic field is broadened mainly by the contribution of $dg\mu B_z/2\pi$, but it is robust against $dE_{1,2}/2\pi$. Hence, $dg\mu B_z/2\pi$ can be estimated from the ODMR width in an external magnetic field. Since the above approach can estimate both $\Gamma/2\pi$ and $dg\mu B_z/2\pi$, $dE_{1,2}/2\pi$ can also be estimated from the ODMR spectrum without an external magnetic field where $\Gamma/2\pi$ and $dg\mu B_z/2\pi$ have already been determined from other simulation results.

The parameters strongly depend on the spin concentration. Figure 4 plots the estimated parameters (obtained from the fitting) as functions of the spin concentration. $dg\mu B_z/2\pi$ (HWHM) (Fig. 4(a)) and $\Gamma/2\pi$ (HWHM) (Fig. 4(b)) nearly linearly depend on the spin concentration, whereas $dE_{1,2}/2\pi$ (HWHM) (Fig. 4(c)) nearly linearly depends on the NV center concentration. The P1 center is considered as an electron spin bath that affects both the inhomogeneous magnetic fields and the high-frequency magnetic field noises. It should be noted that not only the strain distribution but also the inhomogeneous electric fields can contribute to $dE_{1,2}/2\pi$. Electron irradiation induces defects, which may lead to the strain distribution [1]. The NV centers have a negative charge and the same amount of the P1 center has a positive charge after donating an electron to the NV centers, which may cause the inhomogeneous electric fields.

Using knowledge about the dependences of the noise parameters on the NV concentrations created by electron irradiation in the HPHT diamond samples, the ODMR spectra can be theoretically simulated as the NV center concentration changes. This is especially important to optimize the NV center concentration in order to realize high-performance quantum devices. As an example, here the case where the NV center is used as a temperature sensor is considered and the optimized NV center concentration is theoretically predicted to maximize the sensitivity. In the estimated temperature sensitivity, an ideal condition where only the NV centers have electron spins in diamond is considered.

In the ideal condition, diamond has equal amounts of P1 centers and NV centers, but the P1

centers have positive charge states. It should be noted that the positive charge state of the P1 center has no electron spin. Additionally, the magnitude of the *z*-direction of the inhomogeneous strain (electric field) distribution is 50 times smaller than the perpendicular direction of the inhomogeneous strain (electric field) distribution.[16] Here, the microwave power is fixed as an experimentally realizable value in our laboratory. That is, $\lambda/2\pi = 0.29$ MHz. The sensing volume is assumed to be around 0.1 μm³, which was determined by the diffraction-limited detection volume of our confocal microscopy. The estimated temperature sensitivity from the ODMR spectrum is given by

$$\eta = \frac{\sigma}{C_{max} C_T} \frac{1}{\sqrt{1/TN}} \frac{1}{\sqrt{OD_{ND}}},$$

where $\sigma$ is the standard deviation of our measurement system at room temperature using sample 4 (when we drive the system with off-resonant MW). $C_{max}$ is the maximum gradient of the estimated ODMR spectrum. $C_T = 78.6$ kHz/K is the temperature dependence parameter of the zero-field splitting. $T$ is the measurement time. $N$ is the proportion of the NV center concentration of the sample to that of the sample 4. $OD_{ND}$ (=6000) is the blocking fluorescent intensity by a neutral density (ND) filter.

To measure the temperature with the NV center via the ODMR spectra, either the sharp dip structure without an applied magnetic field or a normal peak structure with an applied magnetic field are suitable. It is worth mentioning that although the latter method is more common in previous experimental demonstrations of diamond-based quantum sensors, the former may be more suitable for some applications due to the large gradient with respect to the small change in the microwave frequency. Here, when the normal peak structure toward the high sensitivity is used, the magnetic fields are assumed to be applied exactly along the [001] direction. Hence, every NV center has the same resonant frequency. This is an important assumption to improve the sensitivity when using the peak structure. Otherwise, only part of the NV centers can be controlled by the resonant microwave pulses for the sensing. (On the other hand, when the dip structure is used, every NV center can be naturally involved in the sensing process without such a careful magnetic field alignment. This is one advantage of the dip structure).

Figure 5 plots the numerically estimated temperature sensitivity with respect to the NV center concentration. The black squares are the sensitivity using the sharp dip without an applied magnetic field. The red squares are the sensitivity using the peak structure observed in the normal ODMR spectrum under an external magnetic field exactly applied along the [001] direction. The blue squares show the results of the normal ODMR spectrum with a magnetic field in an arbitrary direction for one of the four NV axes. (It should be noted that each ensemble has four NV axes. Thus, in the case of a magnetic field in an arbitrary direction, four resonance frequencies are present.) The approach using the sharp dip structure has about a three-fold better temperature sensitivity than the approach using the peak structure under an external magnetic field exactly applied along the [001] direction. The estimated optimal sensitivity is around 0.76 mK/$\sqrt{\text{Hz}}$ at room temperature with an NV center concentration of $5.0 \times 10^{17}$/cm$^3$. Our theoretical estimation of the sensitivity is almost one order of magnitude better than the currently reported value (5 mK/$\sqrt{\text{Hz}}$), which was achieved using a confocal microscope with an NV center [12, 13].

Although there are many ways to optimize the performance of the sensors, we especially focus on the concentration of the NV centers for the optimization. Increasing the sensing volume $V$ in the bulk diamond, which is equivalent to an increase in the number of the measured NV centers, can improve the sensitivity by a factor of $\sqrt{V}$. If the sensing volume is assumed to be around 1000 μm$^3$, the estimated temperature sensitivity achieved to ~20 mK/$\sqrt{\text{Hz}}$. Increasing the averaging time and/or decreasing the measurement standard deviation (possibly by employing magnetic shielding) should also help to improve the sensitivity. Furthermore, it should be noted that our estimation considered a relatively low microwave power (below 10 mG) in the ODMR signal. Such low-power microwave driving is important when measuring biomaterials in order to avoid heating the water. Additionally, such a low-power drive prevents heating the target sample (the object whose temperature is measured), which is one of the reasons for the deteriorated performance of the previous temperature sensor [13]. In principle, our optimization is useful not only for the sensing at a room temperature bust also at a high temperature. However, in the case of high temperature (400~700 K), the ODMR contrast and fluorescence intensity decrease [32], and so our estimated sensitivity become lower. Therefore, our estimated sensitivity discussed above can be achieved at around room temperature.

## V. CONCLUSION

We investigated the dependence of the noise parameters of high-density NV center ensembles. Inhomogeneous magnetic fields and homogeneous broadenings nearly linearly depend on the spin concentrations, whereas the inhomogeneous strain (electric field) distribution nearly linearly depends on the NV center concentration. Additionally, to illustrate the importance of such parameter dependences when optimizing a quantum device, we theoretically estimated the influence of spin concentration on the performance of the NV center ensemble as a temperature sensor. Based on the theoretical calculations, the sharp dip structure in the ODMR spectrum is suitable for a temperature sensor. The optimal sensitivity is predicted to occur around 0.76 mK/$\sqrt{\text{Hz}}$ with an NV center concentration of $5.0 \times 10^{17}$/cm$^3$ and a spatial resolution of ~0.1 μm$^3$, which is the diffraction-limited volume in a typical confocal microscope. This estimated value is better than the previous experimental results [12, 13]. Our results are essential to control and use NV center ensembles as high-performance quantum devices, particularly as temperature sensors.

**ACKNOWLEDGMENTS**

This work is supported by KAKENHI (No. 15H05868, 15H05870, 15K17732) and CREST

(JPMJCR1333).

**REFERENCES**

[1] G. Davies and M. Hamer, Proc. R. Soc. Lond. Ser. A Math. Phys. Eng. Sci. **348**, 285 (1976).

[2] A. Gruber, A. Dr¨abenstedt, C. Tietz, L. Fleury, J. Wrachtrup, and C. V. Borczyskowski, Science **276**, 2012 (1997).

[3] F. Jelezko, T. Gaebel, I. Popa, A. Gruber, and J. Wrachtrup, Phys. Rev. Lett. **92**, 076401 (2004).

[4] M. W. Doherty, N. B. Manson, P. Delaney, F. Jelezko, J. Wrachtrup, and L. C. L. Hollenberg, Phys. Rep. **528**, 1 (2013).

[5] T. Shimo-Oka, H. Kato, S. Yamasaki, F. Jelezko, S. Miwa, Y. Suzuki, and N. Mizuochi, Appl. Phys. Lett. **106**, 153103 (2015).

[6] P. Neumann, N. Mizuochi, F. Rempp, P. Hemmer, H. Watanabe, S. Yamasaki, V. Jacques, T. Gaebel, F. Jelezko, and J. Wrachtrup, Science **320**, 1326 (2008).

[7] T. Wolf, P. Neumann, K. Nakamura, H. Sumiya, T. Ohsima, J. Isoya, and J. Wrachtrup, Phys. Rev. X **5**, 041001 (2015).

[8] N. Aslam, M. Pfender, P. Neumann, R. Reuter, A. Zappe, F. F. D. Olivelea, A. Denisenko, H. Sumiya, S. Onoda, J. Isoya, and J. Weachtrup, Science **10**, 1126 (2017).

[9] J. M. Boss, K. S. Cujia, J. Zopes, and C. L. Defen, science **356**, 837 (2017).

[10] S. Schmitt, T. Gefen, F. M. Sturner, T. Unden, G. Wolff, C. Muller, J. Scheuer, B. Naydenov, M. Markham, S. Pezzagna, J. Meijer, I. Sehwarz, M. Plenio, A. Retzker, L. P. McGuinness, and F. Jelezko, science **26**, 832 (2017).

[11] H. Clevenson, M. E. Trusheim, C. Teale, T. Schroder, D. Braje, and D. Englund, Nat. Phys. **11**, 393 (2015).

[12] P. Neumann, I. Jakobi, F. Dolde, C. Burk, R. Reuter, G. Waldherr, J. Honert, T. Wolf, A. Brunner, J. H. Shim, D. Suter, H. Sumiya, J. Isoya, and J. Wrachtrup, Nano Lett. **13**, 2738 (2013).

[13] G. Kucsko, P. C. Maurer, N. Y. Yao, M. Kubo, H. J. Noh, P. K. Lo, H. Park, and M. D. Lukin, Nature **500**, 12373 (2013).

[14] T. Plakhotnik, M. W. Doherty, J. H. Cole, R. Chapman, and N. B. Manson, Nano Lett. **14**, 4990 (2014).

[15] M. W. Doherty, V. V. Struzhkin, D. A. Simpson, L. P. McGuinness, Y. Meng, A. Stacey, T. J. Karle, R. J. Hemley, N. Manson, L. C. C. Hollenberg, and S. Prawer, Phys. Rev. Lett. **112**, 047601(2014).

[16] F. Dolde, H. Fedder, M. W. Doherty, T. Nobauer, F. Rempp, G. Balasubramanian, T. Wolf, F. Reinhard, L. C. Hollenberg, F. Jelezko, and J. Wrachtrup, Nat. Phys. **7**, 459 (2011).

[17] F. Dolde, M. W. Doherty, J. Michi, I. Jalobi, B. Naydenov, S. Pezzagna, J. Meijer, P. Neumann, F. Jelezko, N. B. Manson, and J. Wrachtrup, Phys. Rev. Lett. **112**, 097603 (2014).

[18] E. Bourgeois, A. Jarmola, P. Siyushev, M. Gulka, J. Hruby, F. Jelezko, D. Budker, and M. Nesladek, Nat. Commun. **6**, 9577 (2015).

[19] G. Balasubramanian, P. Neumann, D. Twitchen, M. Markham, R. Kolesov, N. Mizuochi, J. Isoya, J. Achard, J. Beck, J. Tissler, V. Jacques, P. R. Hemmer, F. Jelezko, and J. Wrachtrup, Nat. Mater. **8**, 383(2009).

[20] N. Mizuochi, P. Neumann, F. Rempp, J. Beck, V. Jacques, P. Siyushev, K. Nakamura, D. Twitchen, H. Watanabe, S. Yamasaki, F. Jelezko, J. Wrachtrup, Phys. Rev. B, **80**, 041201(R) (2009).

[21] T. Yamamoto, T. Umeda, K. Watanabe, S. Onoda, M. L. Markham, D. J. Twitchen, B. Naydenov, L. P. McGuinness, T. Teraji, S. Koizumi, F. Dolde, H. Fedder, J. Honert, J. Wrachtrup, T. Ohshima, F. Jelezko, J. Isoya, Phys. Rev. B **88**, 075206 (2013)

[22] X. Zhu, Y. Matsuzaki, R. Amsuss, K. Kakuyanagi, T. Shimo-oka, N. Mizuochi, K. Nemoto, K. Semba, W. J. Munro, and S. Saito, Nat. Commun. **5**, 3424 (2014).

[23] Y. Matsuzaki, X. Zhu, K. Kakuyanagi, H. Toida, T. Shimo-oka, N. Mizuochi, K. Nemoto, K. Semba, W. J. Munro, H. Yamaguchi, and S. Saito, Phys. Rev. A **91**, 042329 (2015).

[24] Y. Kubo, C. Grezes, A. Dewes, T. Umeda, J. Isoya, H. Sumiya, N. Morishita, H. Abe, S. Onoda, T. Ohshima, V. Jacques, A. Dréau, J.-F. Roch, I. Diniz, A. Auffeves, D. Vion, D. Esteve, and P. Bertet, Phys. Rev. Lett. **107**, 220501 (2011).

[25] S. Choi, J. Choi, R. Landif, G. Kucsko, H. Zhou, J. Isoya, F. Jelezko, S. Onoda, H. Sumida, V. Khemani, C. V. keyserlingk, N. Y. Yao, E. Demler, and M. D. Lukin, nature **543**, 221 (2017).

[26] J. Choi, S. Choi, G. Kucsko, P. C. Maurer, B. J. Shields, H. Sumiya, S. Onoda, J. Isoya, E. Demler, F. Jelezko, N. Y. Yao, and M. D. Lukin, Phys. Rev. Lett. **118**, 093601 (2017).

[27] G. Kucsko, S. Choi, J. Choi, P. C. Maurer, H. Zhou, R. Landig, H. Sumiya, S. Onoda, J. Isoya, F. Jelezko, E. Demler, N. Y. Yao, add M. D. Lukin, arXiv:1609. 08216 (2016).

[28] Y. Matsuzaki, H. Morishita, T. Shimo-oka, T. Tashima, K. Kakuyanagi, K. Semba, W. J. Munro, H. Yamaguchi, N. Mizuochi, and S. Saito, J. Phys.: Condens. Matter **28**, 275302 (2016).

[29] F. T. Charnock, and T. A. Kennedy, Phys. Rev. B **64**, 041201R (2001)

[30] R. Hanson, V. V. Dobrovitski, A. E. Feiguin, O. Gywat, and D. D. Awshalom, Science **320**, 352 (2008)

[31] G. D. Lange, Z. H. Wang, D. Riste, V. V. Dobrovitski, and R. Hanson, Science **330**, 60 (2010)

[32] D. M. Toyli, D. J. Christle, A. Alkauskas, B. B. Buckley, C. G. Van de Walle, and D. D. Awshalom, Phys. Rev. X **2**, 031001 (2012).

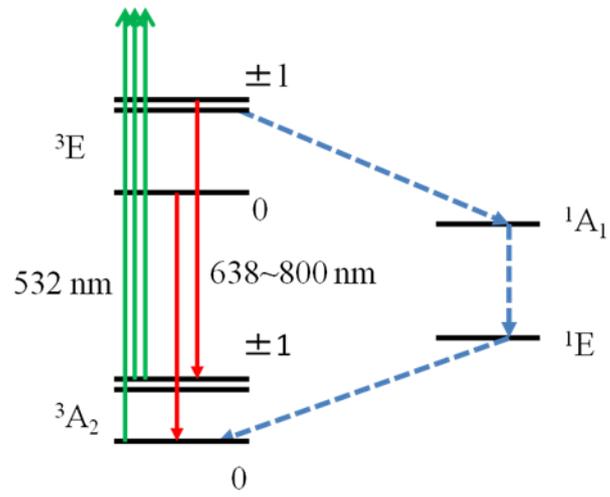

FIG. 1. Energy level diagram of an NV center. $^3A_2$ is the spin-triplet ground states and $^3E$ is the excited states. $^1A_1$ and $^1E$ are the singlet states. Blue dashed line denotes the non-radiative decay path.

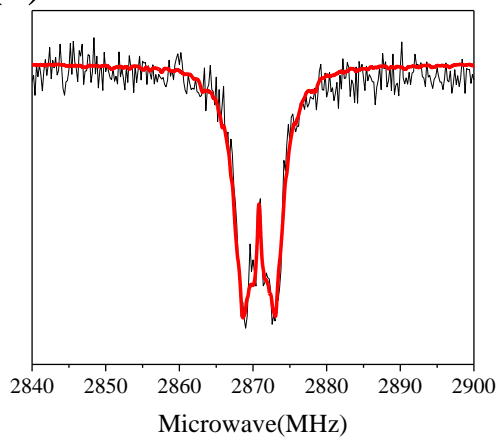
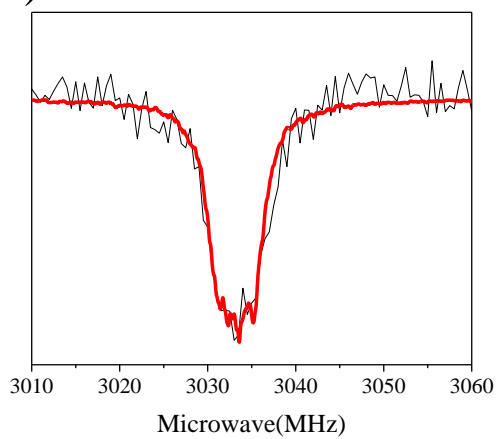
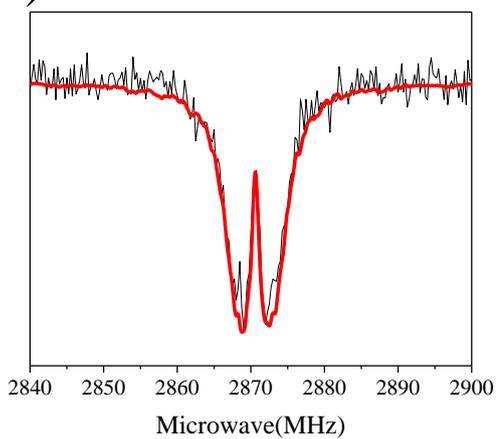
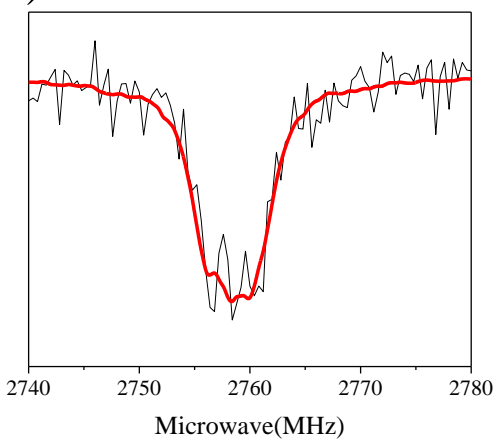
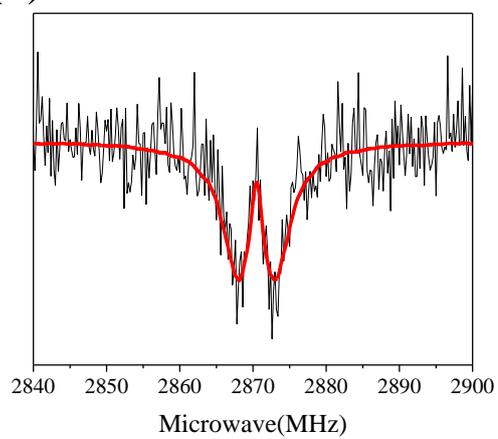
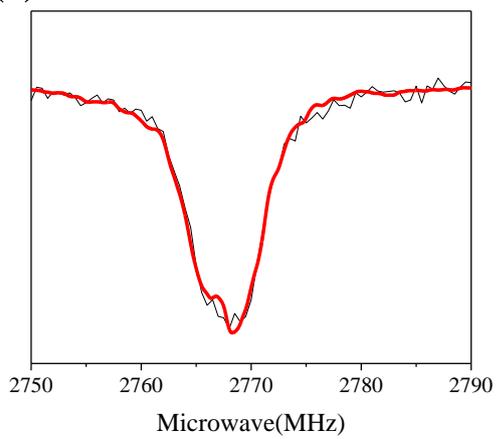

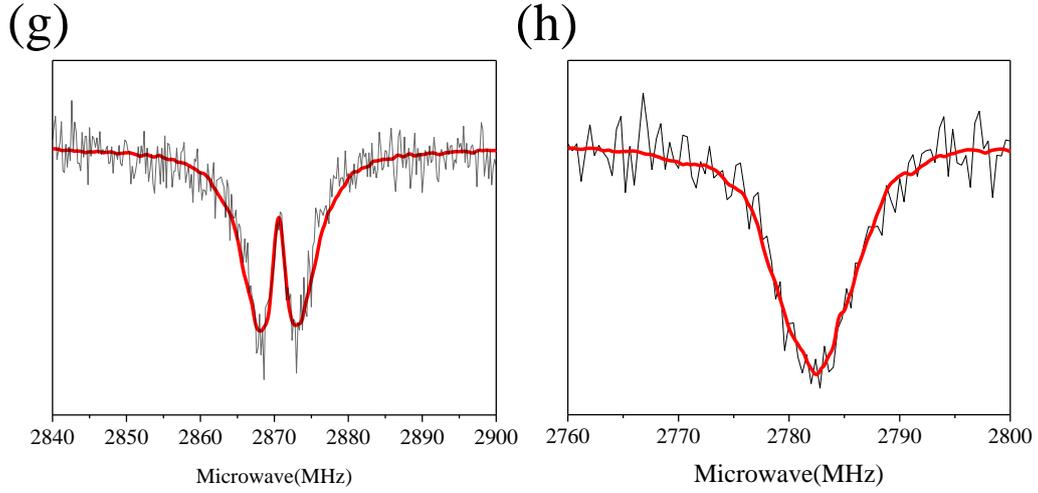

FIG. 2. ODMR simulation results for (a), (b) sample 1, (c), (d) sample 2, (e), (f) sample 3, and (g), (h) sample 4. External magnetic fields are not applied for (a), (c), (e), (g) while we apply the external magnetic fields for (b), (d), (f), (h). Black (red) line is the experimental (numerical) result. Without the external magnetic field, a sharp dip structure is clearly observed around 2.87 GHz. On the other hand, with the external magnetic field, only the normal peak structure is observed around resonant frequencies.

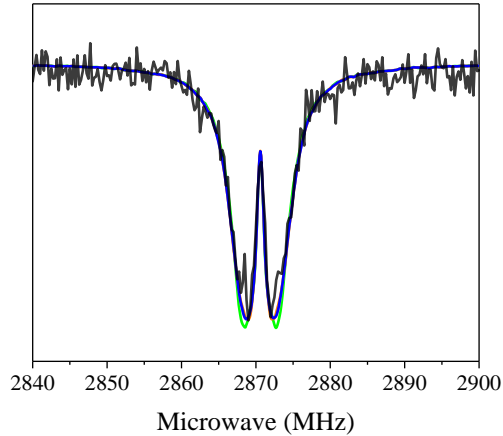

FIG. 3. The simulation results including the earth magnetic field. We assume that the amplitude of the earth magnetic fields is 0.045 mT. The black line showed the experimental result of sample 2. Green line showed the simulation result without the earth magnetic field. Gray (red) line showed the simulation result with the earth magnetic field which direction is (111) ((100)). Blue line showed the simulation result with the earth magnetic field which direction is $\theta=30°\phi=60°$ of polar coordinate.

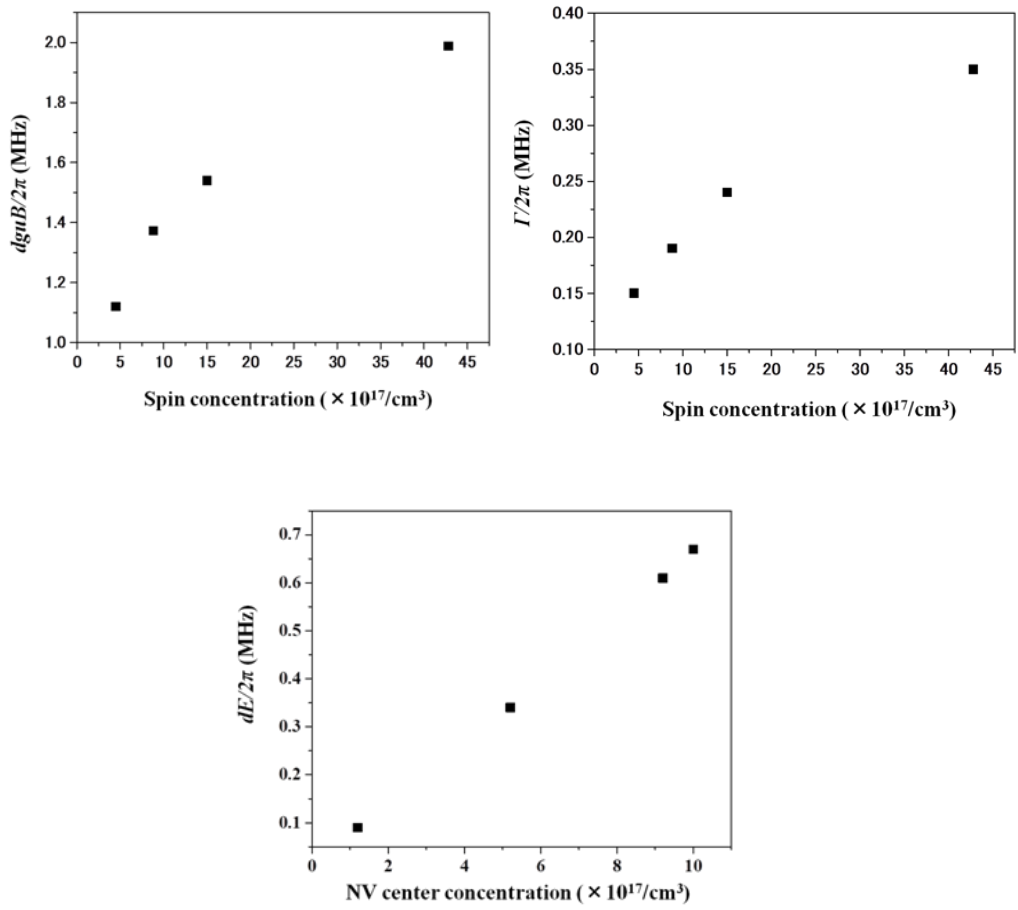

FIG. 4. Plot of (a) dguB/2π(HWHM), (b) Γ/2π (HWHM), and (c) dE/2π (HWHM) with respect to spin (NV center) concentration. Parameters are estimated by fitting the numerical simulations to the experimental results.

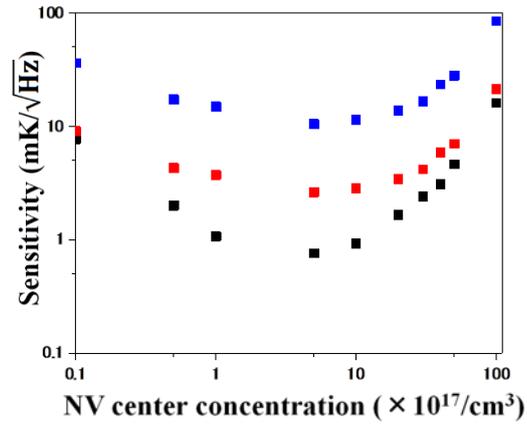

FIG. 5. Numerically estimated sensitivity with respect to the NV concentration with a low microwave power ODMR. Black squares are the sensitivity of the approach using the sharp dip without an applied magnetic field. Red squares are the sensitivity using the peak structure observed in the normal ODMR under an external magnetic field exactly applied along the [001] direction. Blue squares show the results using the peak structure observed in the normal ODMR with an arbitrary direction of the magnetic field and the ODMR signal of only one of four NV axes is measured.

| Sample | P1 center concentration ($\times 10^{17}/cm^3$) | NV center concentration ($\times 10^{17}/cm^3$) |
|---|---|---|
| 1 | 3.3 | 1.2 |
| 2 | 3.6 | 5.2 |
| 3 | 5 | 10 |
| 4 | 33.6 | 9.2 |

Table 1. P1 center concentration and NV center concentration of the four HPHT samples.